\documentclass[prb,aps,floatfix,twocolumn,nopacs,superscriptaddress]{revtex4}

\usepackage[utf8]{inputenc}
\usepackage[american,]{babel}
\usepackage[T1]{fontenc}
\usepackage[pdftex]{graphicx}  
\usepackage{graphicx, xcolor}
\usepackage{dcolumn}
\usepackage{bm}
\usepackage{amsmath,amsthm,amssymb}
\usepackage{hyperref}
\usepackage{xcolor}
\hypersetup{colorlinks,bookmarksopen,bookmarksnumbered,
citecolor=cyan,
linkcolor=cyan,
pdfstartview=false,
urlcolor=cyan}
\usepackage{graphicx}
\usepackage{braket}
\usepackage{wrapfig}
\usepackage{soul}
\usepackage{mathtools}
\usepackage{float}

\renewcommand{\vec}{\boldsymbol}

\begin{document}

\title{Measurement-Induced Entanglement Transitions in the Quantum Ising Chain: \\ From Infinite to Zero Clicks}

\author{Xhek Turkeshi}
\email{xturkesh@sissa.it}
\affiliation{SISSA, via Bonomea 265, 34136 Trieste, Italy}
\affiliation{ICTP, strada Costiera 11, 34151 Trieste, Italy}
\affiliation{JEIP, USR 3573 CNRS, Coll\`{e}ge de France, PSL Research University, 11 Place Marcelin Berthelot, 75321 Paris Cedex 05, France}
\author{Alberto Biella}
\affiliation{Universit\'e Paris-Saclay, CNRS, LPTMS, 91405 Orsay, France}
\affiliation{INO-CNR BEC Center and Dipartimento di Fisica, Universit\`{a} di Trento, 38123 Povo, Italy}
\affiliation{JEIP, USR 3573 CNRS, Coll\`{e}ge de France, PSL Research University, 11 Place Marcelin Berthelot, 75321 Paris Cedex 05, France}
\author{Rosario Fazio}
\affiliation{ICTP, strada Costiera 11, 34151 Trieste, Italy}
\affiliation{Dipartimento di Fisica, Universit\'a di Napoli "Federico II", Monte S. Angelo, I-80126 Napoli, Italy}
\author{Marcello Dalmonte}
\affiliation{SISSA, via Bonomea 265, 34136 Trieste, Italy}
\affiliation{ICTP, strada Costiera 11, 34151 Trieste, Italy}
\author{Marco Schir\'o}
\affiliation{JEIP, USR 3573 CNRS, Coll\`{e}ge de France, PSL Research University, 11 Place Marcelin Berthelot, 75321 Paris Cedex 05, France}

\begin{abstract}
We investigate measurement-induced phase transitions in the Quantum Ising chain coupled to a monitoring environment. We compare two different limits of the measurement problem, the stochastic quantum-state diffusion protocol corresponding to infinite small jumps per unit of time and the no-click limit, corresponding to post-selection and described by a non-Hermitian Hamiltonian. 
In both cases we find a remarkably similar phenomenology as the measurement strength $\gamma$ is increased, namely a sharp transition from a critical phase with logarithmic scaling of the entanglement to an area-law phase, which occurs at the same value of the measurement rate in the two protocols. 
An effective central charge, extracted from the logarithmic scaling of the entanglement, vanishes continuously at the common transition point, although with different critical behavior possibly suggesting different universality classes for the two protocols.
We interpret the central charge mismatch near the transition  in terms of noise-induced disentanglement, as suggested by the entanglement statistics which displays emergent bimodality upon approaching the critical point. The non-Hermitian Hamiltonian and its associated subradiance spectral transition provide a natural framework to understand both the extended critical phase, emerging here for a model which lacks any continuous symmetry, and the entanglement transition into the area law. 
\end{abstract}

\date{\today}
\maketitle

\section{Introduction}

Recent years have seen major progress in the understanding of many-body quantum dynamics. Two well separated limits have been discussed extensively, the unitary dynamics of closed isolated systems~\cite{polkovnikov2011nonequilibrium,dalessio2016fromquantum} and the dissipative dynamics of open quantum systems coupled to an external environment~\cite{breuer2002the}. In the former case the system remains in a pure many-body state which, in absence of ergodicity breaking~\cite{nandkishore2015manybody,abanin2019colloquium}, acts as an efficient bath for any of its subsystem, leading to thermalization of local observables and volume-law entanglement entropy. In the latter, the system is intrinsically mixed and described by a master equation for the density matrix. The competition between unitary and dissipative couplings in this setting can lead to non-equilibrium phase transitions but it is not expected to change the entanglement properties of the system.

An intermediate situation which has recently attracted major attention is the one in which the external environment represents a measurement apparatus, which ceaselessly probe some property of the system. Here the key physics is encoded in stochastic quantum many-body trajectories, which contain a much richer information on the system dynamics than the average state. As a result new dynamical phases arise, which are characterized by different entanglement properties. For generic unitary evolution, as that encoded in random circuits~\cite{li2018quantum,gullans2020dynamical,gullans2020scalable,li2019measurement,li2019conformal,chan2019unitary,jian2020measurement,zabalo2020critical,piqueres2020mean,skinner2019measurement,nahum2021measurement,choi2020quantum,bao2020theory,li2020statistical,fan2020selforganized,szyniszewski2019entanglement,zhang2020nonuniversal,snizhko2020quantum,shtanko2020classical,lunt2021dimensional,turkeshi2020measurement,jian2020criticality,sang2020entanglement,shi2020entanglement,ippoliti2021postselection,lavasani2021measurement,lavasani2020topological,sang2020measurement,ippoliti2021entanglement,lang2020entanglement,turkeshi2021measurement}, these systems show a transition between an error correcting phase and a Zeno phase, signaled by the transition of entanglement from a volume to an area law. 

If instead the unitary dynamics is generated by an Hamiltonian, the properties of the system depend on the nature of the latter~\cite{fuji2020measurement,rossini2020measurement,lunt2020measurement,goto2020measurement,tang2020measurement}. For instance, in the case of a free fermion Hamiltonian, the volume law is unstable for any measurement rate to a subextensive entanglement content~\cite{cao2019entanglement,fiskowski2021how}.  However, it has been shown that the average entanglement entropy can still show a transition between a logarithmic and area law phase at a critical measurement strength or to display a purely logarithmic scaling, depending on the stochastic protocol~\cite{alberton2020trajectory}.  A logarithmic growth of the entanglement entropy in an entire phase is particularly intriguing, given that the average state is expected to be effectively thermal, and it is reminiscent of a critical, conformally-invariant, phase whose origin has been so far elusive. Similar results have been obtained for free-fermion random circuits with temporal randomness~\cite{chen2020emergent},  a setting that has been recently generalized to higher dimension~\cite{tang2021quantum}, or for Majorana random circuits~\cite{nahum2020entanglement,bao2021symmetry}. 

A different take on measurement-induced transitions has instead focused on the limit of post-selection, also called forced measurement phase transitions~\cite{nahum2021measurement}, where only atypical trajectories with a particular outcome of measurement are retained. In this regime several works have discussed the relationship between measurement-induced transitions and the spectral properties of the associated non-Hermitian Hamiltonian~\cite{biella2020manybody,gopalakrishnan2020entanglement,jian2021yang}. More recently a non-Hermitian Hamiltonian has been also shown to emerge from the replica field theory associated to monitored Luttinger Liquids, leading to a Kosterlitz-Thouless transition between a gapless and a gapped phase~\cite{buchhold2021effective}.

In light of these developments it is interesting to compare the entanglement properties of stochastic quantum dynamics and non-Hermitian Hamiltonian, in a simple and paradigmatic setting. With such motivation, in this work we consider the one-dimensional Quantum Ising model coupled to an environment which continuously measures its transverse magnetization. Specifically, we focus on two rather opposite limits of the measurement problem: the quantum state diffusion protocol~\cite{gisin1992the,brun2002a}, a very clear, experimentally realizable measurement protocol in quantum optical systems~\cite{gardiner2000quantum}, equivalent to homodyne trajectories~\cite{wiseman1993quantum,yang2018theory}  and the quantum jump protocol in the so called no-click limit, corresponding to a purely deterministic non-Hermitian evolution, which amounts to post-select only trajectories without jumps.

Remarkably, despite describing rather different limits, show a very similar phenomenology in their entanglement properties. In particular, upon increasing the ratio $\gamma$ between measurement rate and Hamiltonian coupling we find in both protocols a transition from a critical phase  to an area law phase, the former characterized by a logarithmic scaling of the entanglement with time and system size.  Right at the transition point $\gamma_c$, which we find numerically to coincide in the two protocols, the prefactor of the logarithmic entanglement contribution, interpreted as an effective central charge, vanishes continuously. 

We note that an extended phase with logarithmic (critical) scaling of the entanglement was found also for other free-fermionic systems under monitoring, for example the XX chain of Ref.~[\onlinecite{alberton2020trajectory}]. Yet this result occurs here for a model, the Quantum Ising chain, which does not conserve particle number but only its parity, and thus lack any obvious continuous symmetry expected in conventional gapless systems. This suggests that the role of microscopic symmetries in the classification of quantum phases of matter, leading to the textbook distinction between gapped and gapless phases in presence of discrete/continuous symmetries, needs to be revisited for quantum many body systems under monitoring, as recently pointed out~\cite{bao2021symmetry}. 

In this respect the no-click limit offers a natural explanation both for the critical entanglement scaling and for its transition towards an area law: these can be naturally understood by looking at the spectral properties of the non-Hermitian Quantum Ising chain which undergoes, right at $\gamma_c$, a subradiance transition from an extended critical phase with gapless decay modes, to a gapped phase which is smoothly connected to the dark state of the measurement operator.

In spite of the important similarities, the two protocols differ for the behavior of the effective central charge as a function of the measurement rate. In particular we find this quantity to be larger in the no-click limit, a signature of a noise-induced disentangling effect that appears clearly in the statistics of the entanglement entropy, which shows emergent bimodality upon approaching the critical point. Our results highlight  the key role played by the non-Hermitian Hamiltonian for the qualitative understanding of entanglement transitions and suggest that different stochastic ensembles provide different universality classes of a common critical phenomenon.

The paper is structured as follows. In Sec.~\ref{sec:Model} we introduce the model and the measurement protocols. Then we discuss the results for the two protocols in Sec.~\ref{sec:entres}, in particular the entanglement dynamics (Sec.~\ref{sec:ent_growth}), the entanglement scaling with system size (Sec.~\ref{sec:ent_scaling}) and the entanglement statistics (Sec.~\ref{sec:ent_stat}). We detail in the Appendix a derivation of the protocol of interest, and a summary of the numerical methods used for the simulations.

\begin{figure}
	\includegraphics[width=\columnwidth]{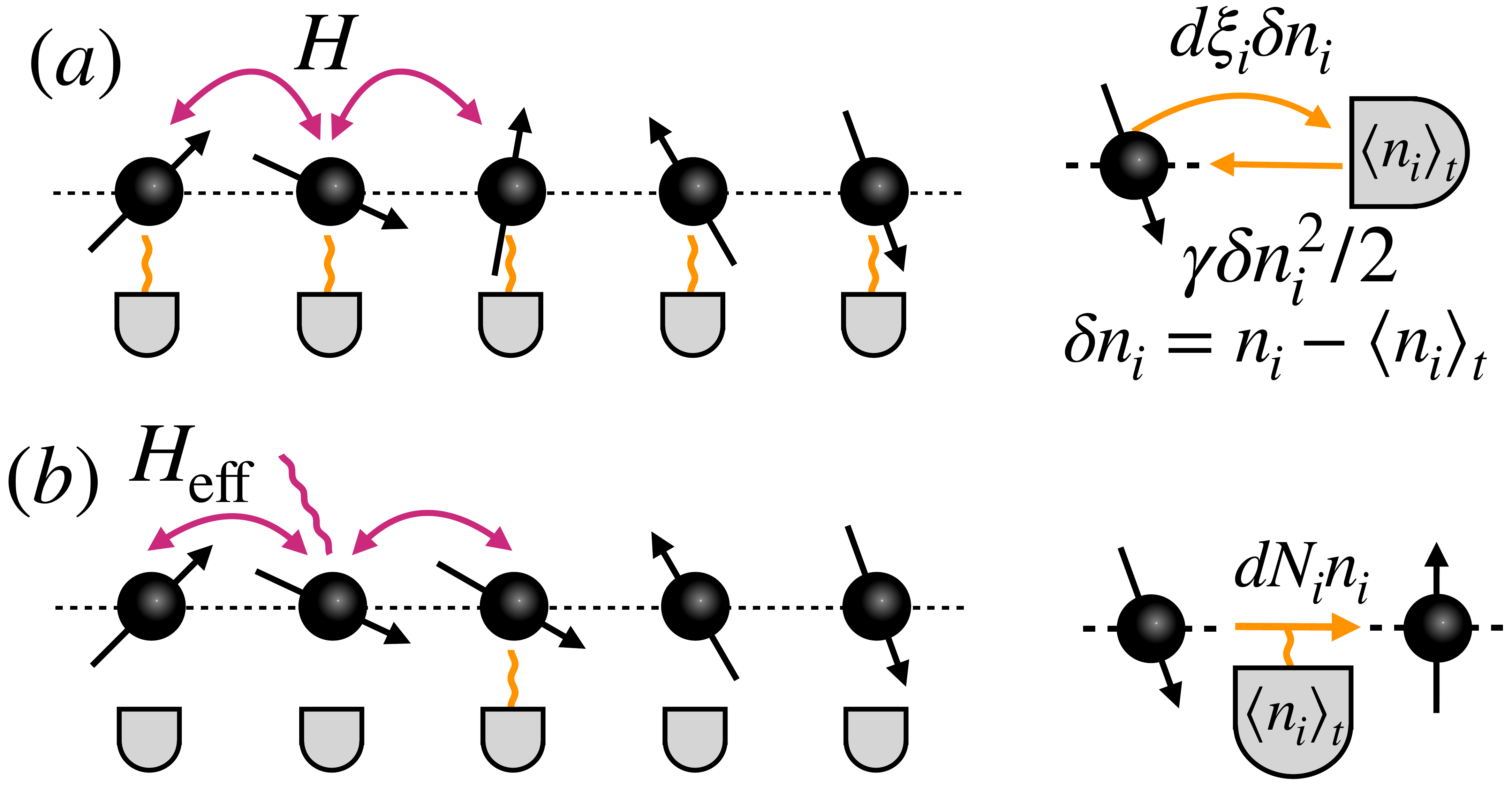}
	\caption{\label{fig:cartoon} Cartoon of the monitored Quantum Ising Chain. (a) Quantum state diffusion protocol: on each lattice site a two-level system (Ising spin) interacts with its neighbors and is subject to a weak-continuous measurement of the up-polarized state. (b) Quantum jump protocol: the spins in the chain interact through a non-Hermitian Hamiltonian. Occasionally a quantum jump take place, projecting the measured degree of freedom in the up-polarized state. The no-click limit corresponds to post-selecting only trajectories without jumps. (See text for details).
	}
\end{figure}

\section{Model and Measurement Protocols}
\label{sec:Model}

We consider a one dimensional Quantum Ising chain with Hamiltonian
\begin{align}
	H = J\sum_{i=1}^{L-1}\sigma^x_i \sigma^x_{i+1}
\end{align}
with open boundary conditions, evolving under the competing effect of its own unitary dynamics and the measurement of the up-polarized state ${n_i\equiv(\sigma^z_i+1)/2=|1\rangle\langle 1|}$ (pictorially illustrated in Fig.~\ref{fig:cartoon}). Here $\sigma^\alpha$ are the Pauli matrices. We now introduce the two measurements protocols that will be considered in this work, namely the quantum state diffusion and the quantum jumps with post-selection, corresponding to the non-Hermitian Hamiltonian.

\subsection{Stochastic Quantum Dynamics with Continuous Monitoring }
\label{sec:model}
Due to the monitoring, the evolution is captured by quantum trajectories $|\psi(\vec{\xi}_t)\rangle$ which follow the quantum state diffusion (QSD) equation
\begin{align}
	d|\psi(\vec{\xi}_t)\rangle &= - i H dt|\psi(\vec{\xi}_t)\rangle  + \sum_{i=1}^L (n_i - \langle n_i\rangle_t)d\xi^i_t  |\psi(\vec{\xi}_t)\rangle \nonumber \\
	&\quad -\frac{\gamma}{2}\sum_{i=1}^L (n_i - \langle n_i\rangle_t)^2 dt |\psi(\vec{\xi}_t)\rangle.
	\label{eq:sse}
\end{align}
(See Appendix~\ref{app:weak} for a brief derivation of the above stochastic Schr\"odinger equation). 
The first term in Eq.~\eqref{eq:sse} represents the unitary evolution, while the remaining encode the noise effects. The $d{\xi}^i_t$ are \^Ito increments of a Wiener process $\vec{\xi}_t=(\xi^1_t,\xi^2_t,\dots,\xi^L_t)$, responsible for the stochastic nature of the trajectories, with zero mean $\overline{d\xi^i_t}=0$ and fulfilling the exact property $d\xi^i_t d\xi^j_t = \gamma dt \delta^{ij}$. 
The last term in Eq.~\eqref{eq:sse} describes a deterministic back-action from the measuring environment. 

We note the presence of a feedback mechanism in Eq.~\eqref{eq:sse}, as the noise contributions couple to the fluctuations of the measured operator $\delta n_i=n_i - \langle n_i\rangle_t$ where $\langle \circ\rangle_t = \langle \psi(\vec{\xi}_t)| \circ|\psi(\vec{\xi}_t)\rangle_t$ is the average over the quantum state. The role of this feedback is to preserve the norm of the state (and any of its cumulant) for any realization of the noise. 

It is important to stress the difference between the conditional and the mean state~\cite{cao2019entanglement}. The conditional state is the quantum trajectory itself $\rho_t(\vec{\xi}_t) = |\psi(\vec{\xi}_t)\rangle\langle \psi(\vec{\xi})_t|$, fixed by a realization of the Wiener process. 
Instead, the mean state is given by
\begin{align}
	\overline{\rho_t} = \int \mathcal{D}\vec{\xi}_t P(\vec{\xi}_t) |\psi(\vec{\xi}_t)\rangle \langle \psi(\vec{\xi}_t)|
	\label{eq:avestate}
\end{align}
with $P(\vec{\xi}_t)$ the probability distribution of the noise. Despite the conditional state is always pure, the mean state is mixed, and follows the  Lindblad master equation
\begin{align}
	\frac{d}{dt}\overline{\rho_t} = - i[H,\overline{\rho_t}] -\frac{\gamma}{2}\sum_i [n_i,[n_i,\overline{\rho_t}]].
	\label{eq:lind}
\end{align}
This difference is crucial, and highlights how the two states $\overline{\rho_t} $ and $\rho_t(\vec{\xi}_t) $ provide rather different information on the statistical properties of the system.
For observables which are linear in the state $O[\rho] = O\rho$, due to the linearity of the disorder average, the mean state $\overline{\rho_t}$ encodes the statistical properties of $O$. Specifically, for any function $f(O)$, we have
\begin{align}
	\overline{\mathrm{tr} (\rho_t(\vec{\xi}_t) f(O))}\equiv \mathrm{tr} (f(O) \overline{\rho_t}).
	\label{eq:4v0}
\end{align}
However, the above relationship fails if one is interested in the noise-average of quantities which depend \emph{non-linearly} on the density matrix. A simple example involve the purity $\mathcal{P}(\rho) = \mathrm{tr}\rho^2$: while for the conditional state $\mathcal{P}(\rho_t(\vec{\xi}_t))=1$ (and hence its disorder average), for the mean state we have $\mathcal{P}(\overline{\rho_t})<1$. 
In general, similar issues arise when one is interested in more subtle statistical correlation of the stochastic process, for example in the so called overlaps $\overline{O_i(t) O_j(t)}=\overline{\langle O_i\rangle_t\,\langle O_j\rangle_t}$. We emphasize that this situation is essentially analogous to the difference between annealed and quenched averages in disordered systems~\cite{frassek2020duality}.

After this parenthesis, we introduce the main observable of interest: the (von Neumann) entanglement entropy~\cite{calabrese2004entanglement,amico2008entanglement,horodeski2009quantum,laflorencie2016quantum}. Given a partition $A\cup B$, the (conditional) reduced density matrix $\rho_A(\vec{\xi}_t) = \mathrm{tr}_B \rho(\vec{\xi}_t)$ encodes the bipartite entanglement. The entanglement entropy is defined as
\begin{align}
	S(\vec{\xi}_t)  = -\mathrm{tr}_A\left[ \rho_A(\vec{\xi}_t)\ln \rho_A(\vec{\xi}_t)\right]. 
	\label{eq:5v0}
\end{align}
We note that this is a well defined measure of entanglement provided that the overall state is pure, in which case it yields the amount of Bell pairs that can be distilled from the quantum state. This is not the case for an overall mixed state, such as the averaged state in Eq.~(\ref{eq:avestate}), where also classical correlation enters into this quantity. For this reason, throughout the paper we only consider the conditional state and the conditional average of the entanglement entropy. To simplify the notation, we furthermore omit the conditional specification.
After evolving the state under Eq.~\eqref{eq:sse}, we compute the entanglement entropy. Its average is given by
\begin{align}
	\overline{S} =  \int \mathcal{D}\vec{\xi}_t P(\vec{\xi}_t) S(\vec{\xi}_t),
\end{align}
as well as its full probability distribution is
\begin{align}
P(S_t)= \int \mathcal{D}\vec{\xi}_t P(\vec{\xi}_t)\;\delta\left(S- S(\vec{\xi}_t)\right)
\label{eq:probaS}
\end{align}
Throughout this paper we set $J=1$ and study the problem by varying the strength of the measurement $\gamma$ and for different system sizes $L$.

\subsection{Non-Hermitian Hamiltonian}
\label{sec:nhh}
When the measurements act occasionally but abruptly on the quantum state, the quantum trajectories are described by the following stochastic Schr\"odinger equation
\begin{align}
	d|\psi(\vec{N_t})\rangle &= -iHdt |\psi(\vec{N_t})\rangle -\frac{\gamma}{2}dt\sum_i \left(n_i-\langle n_i\rangle_t \right)|\psi(\vec{N_t})\rangle \nonumber \\
	&\quad + \sum_i\left(\frac{n_i}{\sqrt{\langle n_i\rangle_t}}-1\right) \delta N^i_t|\psi(\vec{N_t})\rangle,
	\label{eq:qjump}
\end{align}
where $\vec{N_t}$ is a Poisson process, with $\delta N^i_t=0,1$, $(\delta N^i_t)^2 = \delta N^i_t$, and $\overline{\delta N^i_t}=\gamma dt \langle n_i\rangle_t$. (A derivation of Eq.~\eqref{eq:qjump} is given in Appendix~\ref{app:weak}). 
The dynamics generated by Eq.~\eqref{eq:qjump} preserves the purity of the state for each trajectory, and gives an alternative unravelling of the average dynamics in Eq.~\eqref{eq:lind}.

When $\delta N^i_t = 0$ on every site, the evolution is driven by the non-Hermitian Hamiltonian
\begin{align}
	H_\textup{eff} = J\sum_{i=1}^{L-1} \sigma^x_i\sigma^x_{i+1} -i\frac{\gamma}{2}\sum_{i=1}^L n_i,
	\label{eq:nonherm}
\end{align}
while when any $\delta N^i_t = 1$, the last term in Eq.~\eqref{eq:qjump} dominates and projects the measured degree of freedom onto the up-polarized state $|1\rangle_i$.
The feedback constant $\propto i \gamma \sum_i\langle n_i\rangle_t$ enters as a renormalization of the wavefunction within the non-Hermitian evolution driven by Eq.~\eqref{eq:nonherm}.
In the following we consider a post-selection of the stochastic dynamics, which arises when, among the experimental runs, we select a portion of trajectories where measurements  do not take place (no-click limit). 
The evolution is therefore deterministic, and driven by the non-Hermitian Hamiltonian in Eq.~\eqref{eq:nonherm}. In particular, the effect of the environment enters the imaginary part of Eq.~\eqref{eq:nonherm}.
Here we notice that this limit capture different aspects of the weak measurement-induced dynamics compared to the quantum state diffusion. In the framework of quantum jumps, the latter arise in the limit of infinite measurement events, and hence correspond to an infinite-click dynamics~\cite{plenio1998the}. 

The dynamics of this non-Hermitian Hamiltonian has been considered previously in Ref.~[\onlinecite{biella2020manybody}], where a subradiant transition has been observed in relationship to a quantum-Zeno effect. Here we study its entanglement properties, namely we evolve the initial density matrix $\rho_0 = |00\dots 0\rangle \langle 0 0  \dots 0|$ according to
\begin{align}
	\rho(t)=\frac{e^{-i H^{\dagger}_{\rm eff}t}\rho_0e^{i H_{\rm eff}t}}{
	\mbox{Tr}\left(\rho_0 e^{i H_{\rm eff}t}e^{-i H^{\dagger}_{\rm eff}t}\right)	
	} \,,
\end{align}
where the denominator take into account the term proportional to $\gamma dt \sum_i\langle n_i\rangle_t$ in Eq.~\eqref{eq:qjump}.
Then, given a partition $A\cup B$, we construct the reduced density matrix $\rho_A(t) = \mathrm{tr}_B \rho(t)$ which encodes the bipartite entanglement. As stressed before, the non-Hermitian Hamiltonian provide a deterministic protocol, hence the entanglement entropy is deterministic
\begin{align}
	S(t)  = -\mathrm{tr}_A \left[\rho_A(t)\ln \rho_A(t)\right]. 
	\label{eq:ee_nonherm}
\end{align}
To explore the stationary regime, in order to avoid residual dynamical oscillations, we consider
\begin{equation}
{S(\infty)}\equiv\lim_{t\to\infty}\frac{1}{t-t_\mathrm{sat}} \int_{t_\mathrm{sat}}^t ds S(s).
	\label{eq:ee_nonherm_infty}
\end{equation}
As for the monitoring case we make use of the Jordan-Wigner mapping to write down the non-Hermitian Ising model in Eq.~\eqref{eq:nonherm} in terms of a quadratic fermionic model (see Appendix~\ref{app:ff}).

\begin{figure}[t!]
	\includegraphics[width=\columnwidth]{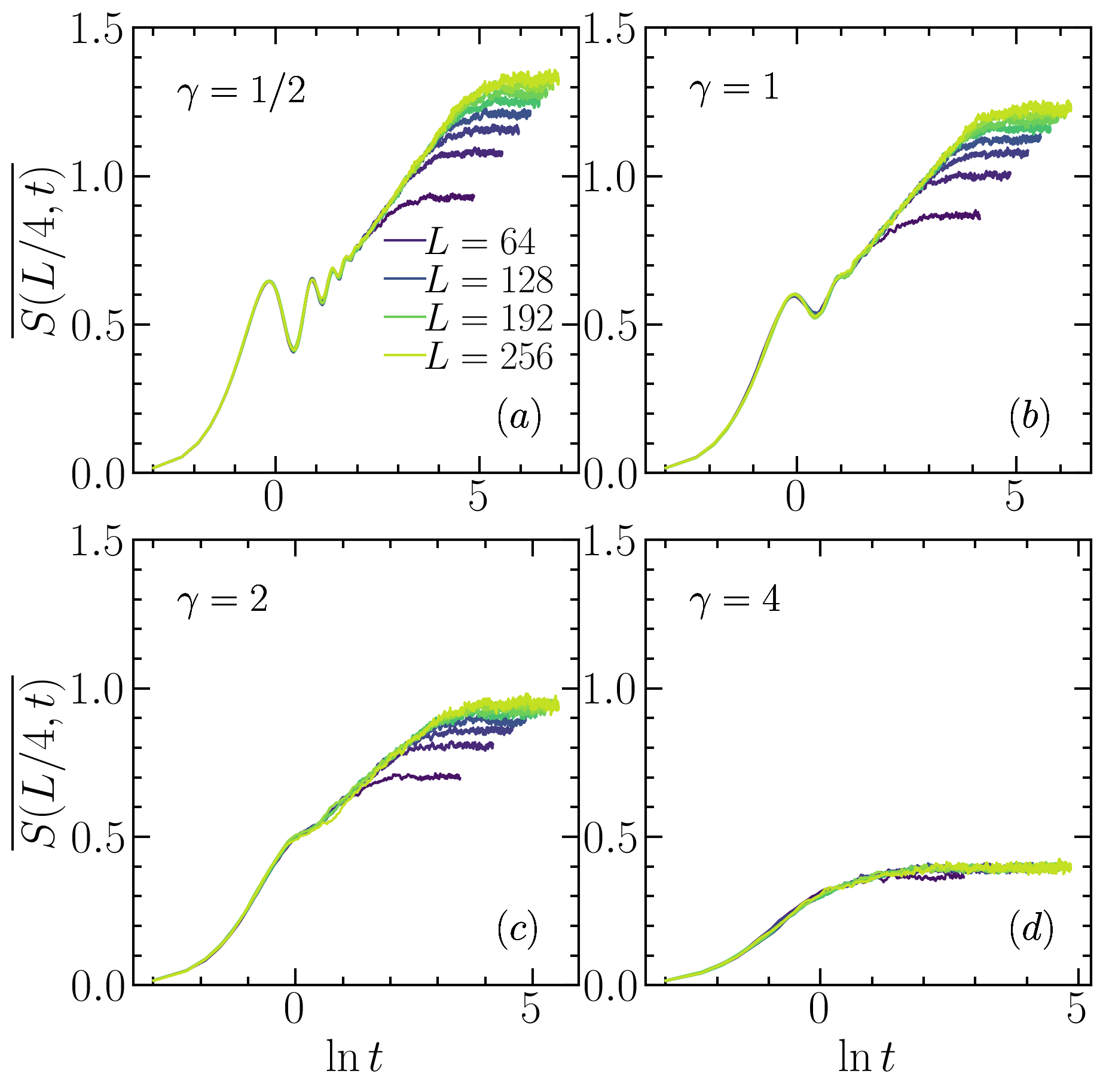}
	\caption{\label{fig:tevo} Dynamics of the entanglement entropy averaged over $N=2000$ trajectories for different values of the measurement strength $\gamma$, different system sizes $L$ and fixing $L_A=L/4$. (a-c) We see a logarithmic growth in time of the entanglement (note the log scale for the time axis) with a saturation to a size-dependent plateau, for $\gamma<4$ and a much slower growth to a value independent from system size, for $\gamma>4$.}
\end{figure}

\section{Results}
\label{sec:entres}
In this section we present the results of the numerical simulations for the quantum state diffusion and the non-Hermitian Hamiltonian. After discussion the entanglement growth within both paradigms, we compare the entanglement scaling at the stationary state. Finally, we investigate the entanglement statistics of the quantum state diffusion protocol, and relate it to the non-Hermitian results.

\subsection{Entanglement Growth}\label{sec:ent_growth}

We start considering the initial state $|\psi_0\rangle = |00\dots 0\rangle$ and evolve the system according to the QSD protocol
defined in Eq.~\eqref{eq:sse}. We make use of the Jordan-Wigner mapping to write the monitored Ising model as a quadratic model of spinless fermions, which can be solved efficiently. (Details on the free fermion techniques, and on the numerical implementation are given in Appendix~\ref{app:ff}). For the computation of entanglement, we choose a subsystem $A$ of $L_A$ contiguous sites.
The average entanglement evolution is shown in Fig.~\ref{fig:tevo}. For small to intermediate values of $\gamma$ (see panels a-c) we find a logarithmic growth in time of the entanglement,
\begin{align}
	\overline{S(t)}\sim \ln t,
\end{align}
which ultimately saturates at times $t_\textup{sat}\sim L/\gamma$. 
It is worth emphasizing that already a small monitoring rate $\gamma$ is enough to give rise to a non-trivial entanglement dynamics compared to the isolated system oscillation (generated by $H=\sum_i \sigma^x_i \sigma^x_{i+1}$). In other words the interplay between measurements and unitary evolution can increase the entanglement production, at least in a certain region of the parameters, as also noted in Ref.~[\onlinecite{biella2020manybody}]. However, upon further increasing the measurement rate we see a substantial change of the entanglement growth which already for $\gamma=4$ shows a rapid saturation to a plateau that is almost independent from the size, a trend that continues for larger values of $\gamma$.

\begin{figure}[t!]
	\includegraphics[width=\columnwidth]{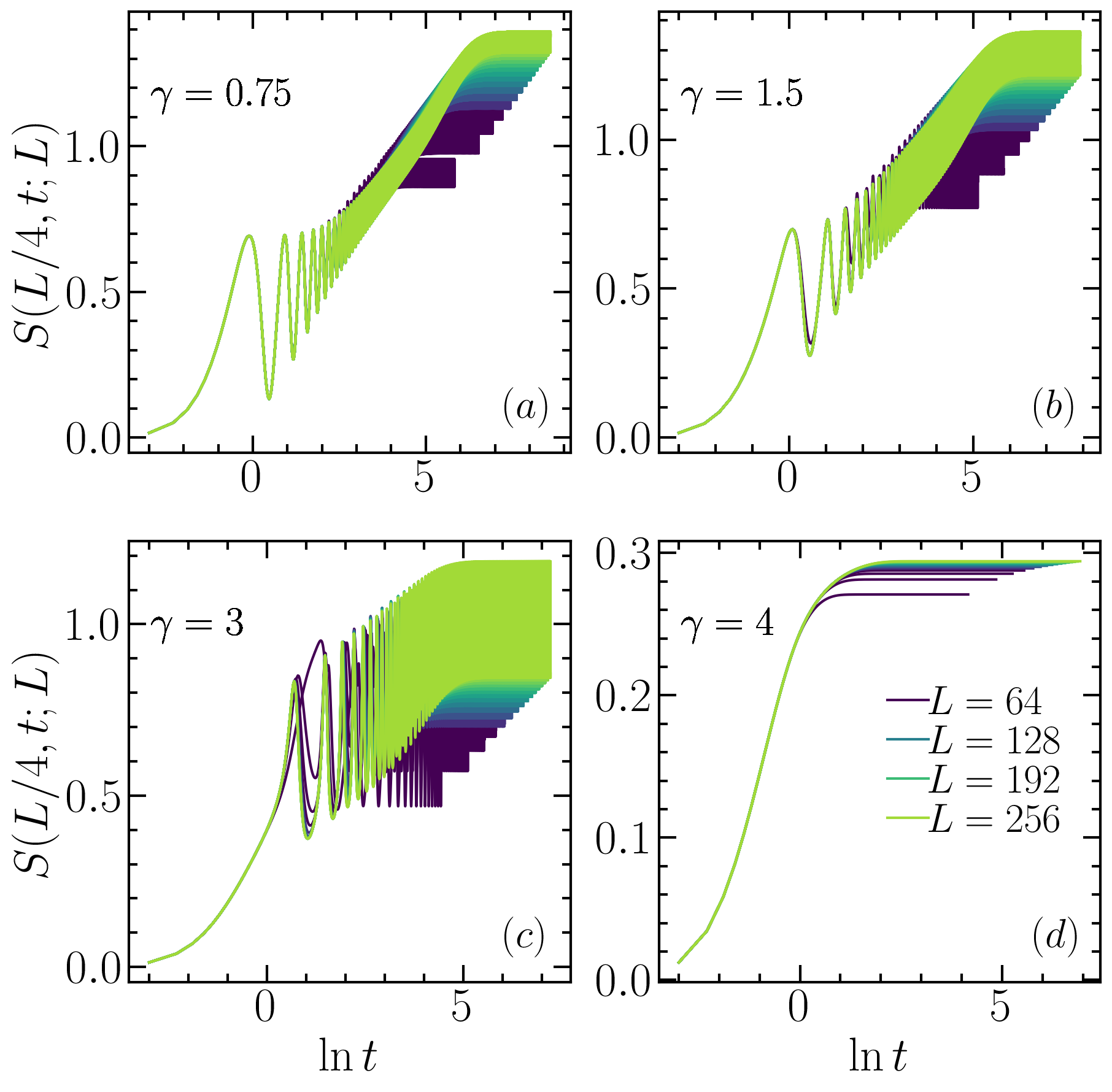}
	\caption{\label{fig:ee_nonherm_tevo} Entanglement dynamics for the non-Hermitian Hamiltonian for different values of the measurement rate $\gamma$ and different system sizes. For small $\gamma$ (panel a) we see a logarithmic growth in time and a saturation to a plateau that depends on system sizes. For large $\gamma$ (panel b) the entanglement rapidly saturates to a value that does not depend on size. Approaching the transition point we see strong oscillations in the dynamics.	}
\end{figure}
The results of the entanglement dynamics for the non-Hermitian Ising model (cfr. Eq.~\eqref{eq:ee_nonherm}) are given in Fig.~\ref{fig:ee_nonherm_tevo}. 
For small values of $\gamma$ (panels a-c), we see that the entropy exhibits an overall logarithmic growth, which resembles the result for the QSD protocol, but has the important difference of exhibiting residual oscillations which dress both the growth and the saturation regimes. Upon further increasing the measuring strength $\gamma$  the entanglement dynamics in the non-Hermitian case displays a sharp transition towards a regime characterized by a fast approach to a stationary value, with no oscillations.  Our numerics locate the boundary between these dynamical regimes around $\gamma\sim 4$, an estimate compatible with what found for the QSD protocol. As we will discuss in the next section we can understand the origin of this sharp transition by looking at the spectral properties of the non-Hermitian Hamiltonian.

\begin{figure}[t!]
	\includegraphics[width=\columnwidth]{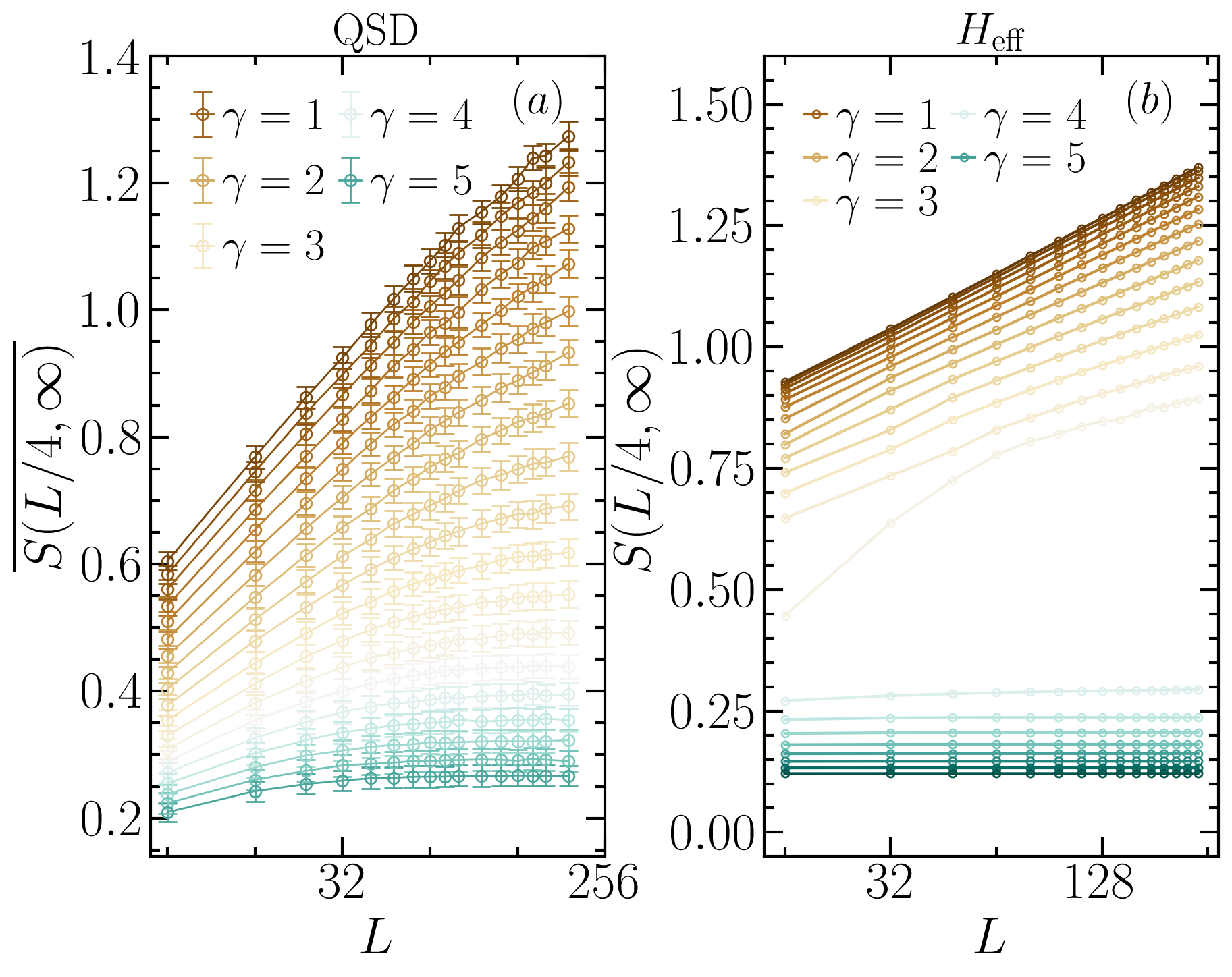}
	\caption{\label{fig:sys} Stationary-state entanglement entropy as a function of the length of the chain. (a) Average entanglement entropy for the QSD protocol. (b) Long-time entanglement entropy for the non-Hermitian dynamics. In both cases we see a transition from a logarithmic \emph{critical} scaling, for small $\gamma$ to an area law (constant) scaling at large $\gamma$. Different curves in the two panels correspond to increasing values of $\gamma$, from $\gamma=0.25$ (top curve) to $\gamma=6$ (bottom curve) with a $\Delta\gamma=0.25$.}
\end{figure}

\subsection{Entanglement Scaling and Effective Central Charge }\label{sec:ent_scaling}

As we have seen, at long enough times the entanglement reaches a stationary value in both protocols. We first discuss the scaling of this stationary entanglement with system size. This usually provides key insights on the properties of the system at long-times. To this extent we chose a subsystem of $L_A=L/4$ contiguous spins and plot in  Fig.~\ref{fig:sys} the long-time entanglement as a function of $L$ and different values of $\gamma$.

For the quantum state diffusion protocol, we can clearly distinguish a regime of logarithmic entanglement growth with system size for small $\gamma$, which eventually evolves into a much weaker dependence, an almost constant plateau for large measurement rates (see Fig.~\ref{fig:sys} (a)). As we are going to discuss further below those two regimes are separated by a sharp entanglement transition.

We note that a logarithmic scaling of the average entanglement in an extended region of parameters is a particularly intriguing result which was obtained before in free-fermion models with a continuous $U(1)$ symmetry~\cite{alberton2020trajectory} and more recently in connection with a BKT transition~\cite{bao2021symmetry,buchhold2021effective}. Here we find this result for a model which only features a discrete $Z_2$ symmetry and whose ground-state would show logarithmic scaling of the entanglement only at the critical point. This suggests that a different mechanism, not directly related to the microscopic symmetries of the problem, could be behind the origin of this seemingly \emph{critical} phase.

Interestingly, the finite-size scaling of the entanglement entropy for the non-Hermitian Hamiltonian shows a behavior which is qualitatively similar to the QSD protocol, as we show  in Fig.~\ref{fig:sys} (b), including a logarithmic scaling for small $\gamma$ and an area-law (constant) scaling for large measurement rates, separated by a sharp transition. We also note that the two scaling regimes are connected by a finite-size crossover, absent in the stochastic QSD protocol, whose characteristic scale we found to diverge at the transition (not shown).

For equilibrium quantum critical systems the prefactor of the logarithmic scaling of the entanglement entropy is controlled by the central charge of the associated Conformal Field Theory(CFT)~\cite{calabrese2004entanglement}, a universal quantity which roughly measures the number of degrees of freedom in the CFT. Following this analogy Ref.~[\onlinecite{alberton2020trajectory}] introduced an
 \emph{effective central-charge} for an XX spin-chain under continuous monitoring, which turned out to be a non-universal function of the measurement rate. Here we follow this route and define an effective central charge for both the QSD protocol and the non-Hermitian Hamiltonian by fitting the data in Fig.~\ref{fig:sys} according to the ansatz
\begin{equation}
\begin{split}
\mathrm{QSD}&\qquad\qquad  \overline{S}= \frac{1}{3}c_\mathrm{QSD}(\gamma)\ln L +a\,,
	\\
H_\mathrm{eff}&\qquad\qquad  S= \frac{1}{3}c_{H_\mathrm{eff}}(\gamma)\ln L +b\,.
\end{split}
\label{eq:S_vs_L}
\end{equation}
Here $a$ and $b$ are two system-size independent quantities. 

\begin{figure}[t!]
	\includegraphics[width=\columnwidth]{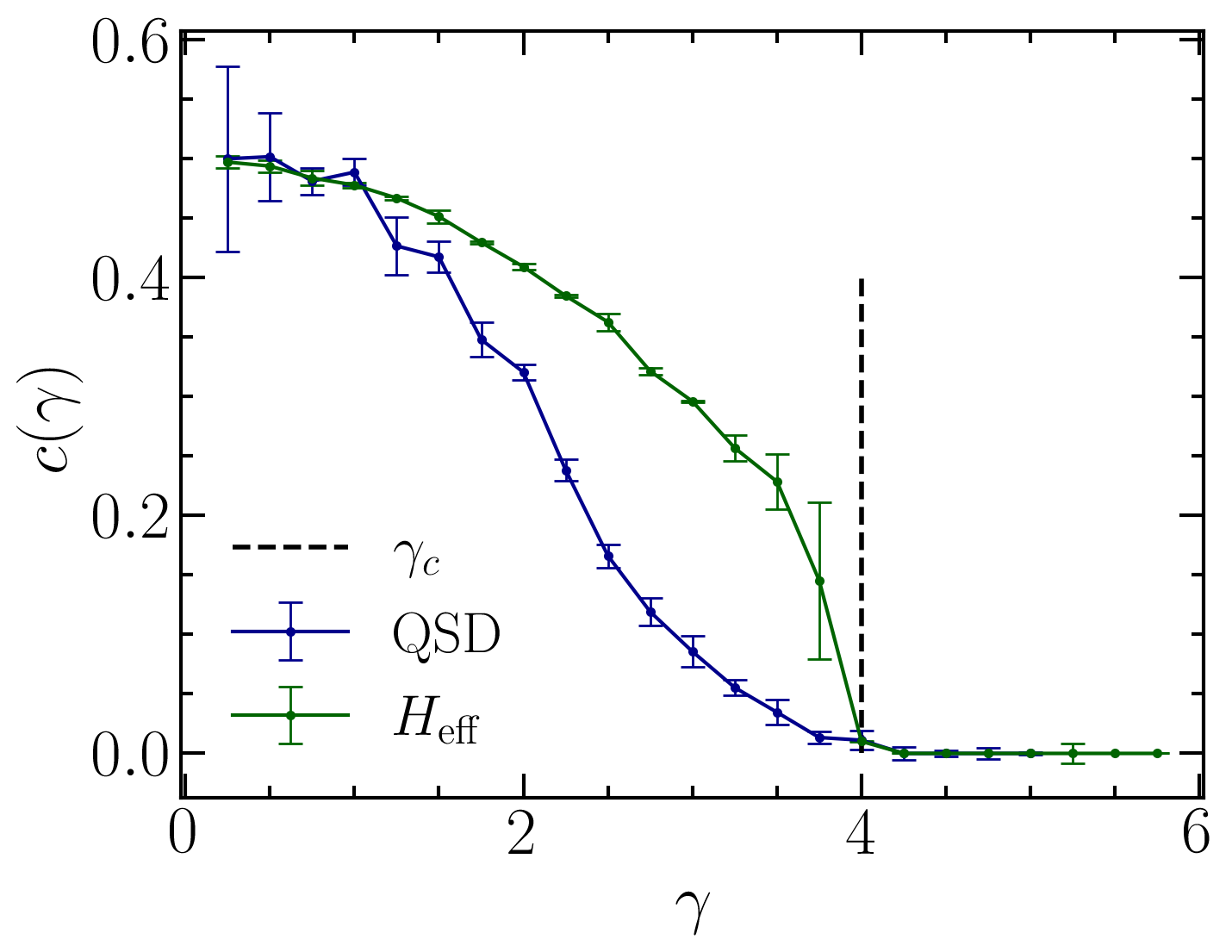}
	\caption{\label{fig:ee_nonherm_scal}  Effective central charge as a function of the measurement rate $\gamma$, for the QSD protocol and the non-Hermitian dynamics. In both cases we find that $c(\gamma)$ decreases monotonously and vanishes around $\gamma_c\simeq 4$, which locates the entanglement transition between logarithmic scaling and area law. Interestingly  the two central charges coincide at small and large values of $\gamma$, while approaching the critical point we find the non-Hermitian one to be larger, possibly vanishing at $\gamma_c$ in a non-analytic way. }
\end{figure}

We plot the result in Fig.~\ref{fig:ee_nonherm_scal} and find that $c(\gamma)$ decreases with $\gamma$ and, quite remarkably, vanishes  for both protocols around the same value of measurement strength $\gamma_c\simeq 4$, that we therefore establish as critical value for the entanglement transitions between logarithmic and area law regimes.
Despite this similarity we notice that the effective central charge shows a rather different dependence from $\gamma$ in the QSD and in the non-Hermitian Hamiltonian protocols. In particular, a direct comparison in Fig.~\ref{fig:ee_nonherm_scal} shows that the effective central charges coincide at small and large values of $\gamma$, while deviate in a region before the critical point. Here the non-Hermitian effective central charge turns out to be larger than the one associated to the QSD protocol, a point on which we will come back in the next section.

Before closing this section we note that many of the properties found for the entanglement dynamics of the non-Hermitian Hamiltonian  can be naturally understood in terms of the spectral properties of the non-Hermitian Ising model in Eq.~(\ref{eq:nonherm}). In fact this can be easily diagonalized once written in fermionic language~\cite{hickey2013time,lee2014heralded,biella2020manybody}, and takes the form
\begin{align}
	H_{\rm eff}=\sum_{k>0}\Lambda_k
	\left(\bar{\eta}_k\eta_k+\eta_{-k}\bar{\eta}_{-k}\right) +E_0,
\end{align}
where $\bar{\eta}_k,\eta_k$ are fermionic operators obtained from a non-Hermitian Bogolyubov rotation, $E_0$ is an additive constant fixing the energy zero, and $\Lambda_k$ are the quasiparticle energies. We note that these are in general complex, as the model lacks any PT symmetry and read
\begin{align}\label{eq:qp_spectrum}
	\Lambda_k=2\sqrt{1-\frac{\gamma^2}{16}+i\frac{\gamma}{2}\cos k}
\end{align}

\begin{figure*}[t!]
	\includegraphics[width=\textwidth]{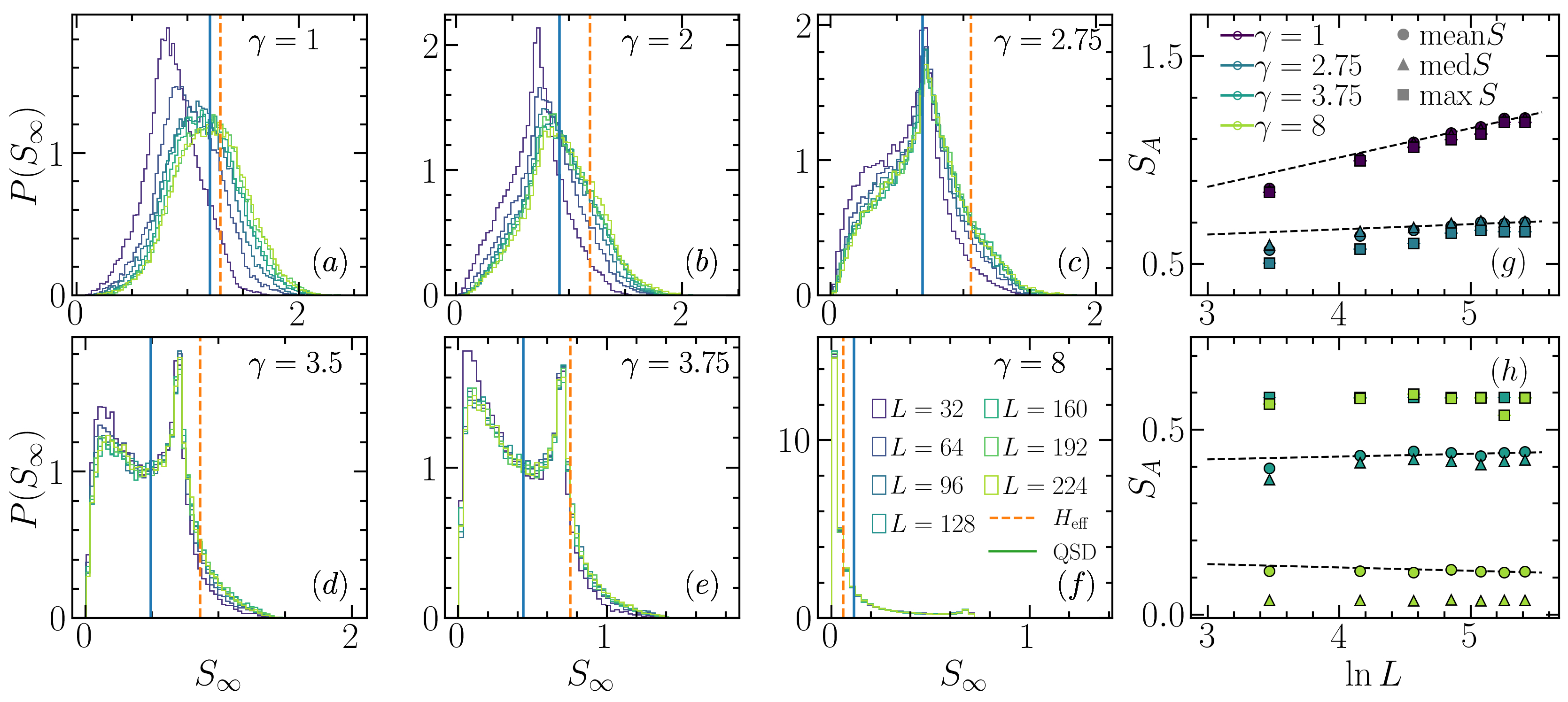}
	\caption{\label{fig:hysto1} (a-f) Probability distribution of the entanglement entropy for different values of $\gamma$ and different system sizes. The realization collected for these functions are $N=64000$. We see that upon increasing $\gamma$, the statistic evolves from a simple Gaussian (panel a-b) to a skewed distribution (panel c-e), with emergent bimodality. At large $\gamma$, the distribution is peaked at $S_\infty\simeq 0$ (panel f). 
	The solid line show the average (blue line) over the trajectories for the quantum state diffusion, and the stationary value of the non-Hermitian Hamiltonian (orange line).
	(g-h) Different statistical estimators $\langle\langle S\rangle\rangle$ versus system size for different values of $\gamma$. Note the logarithmic scaling in the $L$-axis.  
	}
\end{figure*}
The non-Hermitian Ising model features a subradiance transition at a critical value of the measurement rate, given exactly by $\gamma_c=4$, at which the state with smallest imaginary part changes in a non-analytic way~\cite{Biella2020}. For small $\gamma<\gamma_c$ the imaginary part of the spectrum in Eq.~(\ref{eq:qp_spectrum}) is \emph{gapless} and, consistently, the correlation functions in this phase decay as a power-law~\cite{lee2014heralded}.  We emphasize that for a non-Hermitian problem the usual Goldstone theorem relating a gapless spectrum with a continuous broken symmetry does not hold. From this point of view the logarithmic scaling of the entanglement entropy can be naturally understood, since the entire phase $\gamma<\gamma_c$ has soft-modes with small quasi-degenerate lifetime, which are responsible for the slow growth of the entanglement. On the other hand the real-part of the spectrum is gapped for $\gamma<\gamma_c$, leading to residual oscillations in the entanglement dynamics.

Above the subradiance transition for $\gamma>\gamma_c$ a gap opens up in the imaginary part of the spectrum $\Lambda_k$, 
which is consistent with the fast entanglement dynamics found in this regime. The correlation functions in this phase are expected to decay exponentially over a length diverging as $\gamma\rightarrow\gamma_c^+$, a fact that can naturally explain the emergence of a crossover in the finite-size entanglement scaling. We note that in the subradiant phase the long-time dynamics is controlled by a pair of  states with the smallest non-zero imaginary part. These in the strong measurement limit coincide with the dark states of the measurement operator, which are factorized (uncorrelated) eigenstates of the transverse magnetization, therefore justifying the small, system-size independent entanglement entropy. 

The insights on the non-Hermitian Hamiltonian could suggest that an effective field theory description could be possible for $\gamma<\gamma_c$ and that the effective central charge could be computed using field theory replica techniques~\cite{calabrese2009entanglement}, which however we do not attempt here. We also note that, in this perspective, an effective central charge which depends from a parameter in a non-universal way seems very natural, and in fact several non-unitary field theories have been shown to enjoy this property~\cite{bianchini2014entanglement,couvreur2017entanglement,chen2020emergent}.

\subsection{Entanglement Statistics}\label{sec:ent_stat}

Additional insights are given by the statistical fluctuations of the entanglement in the quantum state diffusion protocol, a feature that goes beyond the simple average so far discussed and that is encoded in the full probability distribution, $P(S_t)$, defined in Eq.~(\ref{eq:probaS}). Concretely, we investigate this distribution at stationarity and evaluate $P(S_\infty)= \lim_{t\to\infty} P(S_t)$.

As we see in Fig.~\ref{fig:hysto1}, the statistics of entanglement shows a rather rich evolution with $\gamma$. In particular we see that in the logarithmic phase ($\gamma<\gamma_c$), the statistics changes dramatically, much more than what one could have guessed from the average value, whose dependence from $\gamma$ can be reabsorded in a renormalization of the central charge. 
For small values of the measurement rate, $\gamma\lesssim 1$ (panel a), the entanglement distribution is a normal Gaussian which broadens asymmetrically for $\gamma\sim 2$ (panel b) and develops an emergent bimodality as the critical point $\gamma_c$ is  approached  (panels c-e). Specifically we see that already for $\gamma=2.75$ a secondary peak, for lower values of the entanglement, starts to emerge. Upon further increasing $\gamma$ we see a transfer of weight between the high-entanglement and low-entanglement peaks, with the latter ultimately becoming the dominant one as the system enters the area-law phase. Here, for $\gamma>\gamma_c$ the entanglement distribution is strongly skewed and peaked around small values of the entropy.

It is interesting to discuss the size dependence of the histograms, which is rather different depending on the value of $\gamma$. In particular, as we show in Fig.~\ref{fig:hysto1} (g), where we plot the average, median and typical entanglement for different $L$ and $\gamma$, we find that for $\gamma\lesssim 2.5$ all indicators scale logarithmically with $L$. Instead, Fig.~\ref{fig:hysto1} (h) shows that approaching the critical point ($\gamma=3.75$) and above it ($\gamma=8$)  the statistical estimators deviate from each others, a signature of the non-gaussianity of the entanglement distribution. In the area-law phase for $\gamma>\gamma_c$ the scaling of all the statistical proxies considered becomes independent on the system size.

The emergence of bimodality in the entanglement statistics is an interesting result, which suggests that within the log-phase the entanglement is not a self-averaging quantity, and that limiting the discussion to the average value could hide important features of the problem. A similar result was found in the monitored XX model under QSD protocol~\cite{alberton2020trajectory} and, in Ref~[\onlinecite{biella2020manybody}] for the statistics of the return probability and the local transverse magnetization under a quantum jump stochastic protocol. In this latter case it was shown that the non-Hermitian Hamiltonian captured a secondary peak in the distribution of the relevant observables.

It is therefore interesting to compare the  statistical properties of the entanglement in the QSD protocol with the non-Hermitian Hamiltonian. This comparison is proposed in Fig.~\ref{fig:hysto1}, where we indicate with a dashed line the long-time entanglement in the non-Hermitian case and with a blue line the average entanglement of the QSD protocol. We see that indeed the non-Hermitian Hamiltonian provides a good proxy for the average entanglement both for small $\gamma$, when entanglement fluctuations are gaussian, as well as for large $\gamma$ when the statistics is dominated by small values of the entanglement entropy. For intermediate values of $\gamma$ instead, when the statistics becomes strongly bimodal and fluctuations due to the noise are more relevant, the entanglement content of the non-Hermitian Hamiltonian is systematically larger than the average, reflecting the atypical nature of the no-click process. This is not surprising in retrospect, as the QSD protocol contains randomness which is expected to decrease the entanglement content of the system as opposed to the translational-invariant non-Hermitian protocol. 
We note also that in addition to the leading scaling behavior of the entanglement, captured by the effective central charge, there is also a non-universal correction to the entanglement that depends in principle on $\gamma$ and could play a role in the precise comparison between the two protocols.

\section{Conclusion}

In this work we have discussed measurement-induced criticality in a Quantum Ising chain coupled to a monitoring environment. We have focused on two different limits of the measurement problem, corresponding respectively to the quantum state diffusion protocol and the no-click dynamics, the latter described by a non-Hermitian Ising model.  

In both cases we found a sharp phase transition in the entanglement properties of the system, as a function of the measurement rate $\gamma$.  The entanglement dynamics out of an initial product state evolves from a slow logarithmic growth at small measurement rates to a fast approach to a stationary value at large $\gamma$. The stationary state entanglement shows a \emph{critical} logarithmic scaling with respect to system size, with an effective central charge which changes continuously with $\gamma$ and vanishes at a critical strength $\gamma_c$, indicating a transition towards an area-law scaling which we found numerically to coincide in the two protocols.
The existence of an extended critical phase with logarithmic scaling of the average entanglement  is particularly surprising for our Ising model which, differently from others stochastic measurement problems considered recently in the literature, lacks a continuous symmetry usually associated with gapless critical behavior. The no-click limit and associated non-Hermitian evolution provides a natural mechanism for this phenomenology, in terms of a spectral (subradiant) phase transition from a critical phase with gapless decay modes at small measurements to a gapped area-law phase.
We also find important differences between the two protocols, in particular in the behavior of the effective central charge as a function of the measurement strength which turns out to be larger in the non-Hermitian case. This effect, which seems to suggest different universality classes depending on the specific measurement ensemble, can be understood in terms of a disentangling effect of the noise.

\begin{acknowledgements}
XT thanks S. Pappalardi, V. Vitale for discussions on related topics. 
We acknowledge computational resources on the Coll\'ege de France IPH cluster. XT acknowledges computing resources at Cineca Supercomputing center through the Italian SuperComputing Resource Allocation via the ISCRA grant EntHybDy.
The work of MD and XT is partly supported by the ERC under grant number 758329 (AGEnTh), by the MIUR Programme FARE (MEPH), and has received funding from the European Union's Horizon 2020 research and innovation programme under grant agreement No 817482. R.F. acknowledges partial financial support from the Google Quantum Research Award. This work has been carried out within the activities of TQT.
AB acknowledges funding by LabEx PALM (ANR-10-LABX-0039-PALM). MS acknowledges support from the ANR grant "NonEQuMat" (ANR-19-CE47-0001). 
\end{acknowledgements}

\appendix
\section{Continuous measurements and stochastic Schr\"odinger equations}
\label{app:weak}
This section is an overview of stochastic quantum dynamics in the Hilbert space (see Ref.~[\onlinecite{plenio1998the,brun2002a,jacobs2006a,wiseman2009quantum}] for general review). We first introduce the positive-operator value measurements (POVM), which arise naturally when system and environment interact for a finite time. In the limit of weak measurement, the system evolve through a stochastic Schr\"odinger equation, the details of which depend on the environment specifics. 
We consider two possible choices of system-environment setups which naturally translate, respectively, in the quantum state diffusion Eq.~\eqref{eq:sse} and in the quantum jump equation Eq.~\eqref{eq:qjump}. From the latter, the non-Hermitian quantum Hamiltonian Eq.~\eqref{eq:nonherm} is obtained through post-selection. 
These equations correspond to different unravelling of the same master equation, highlighting the independence of the mean state Eq.~\eqref{eq:avestate} from the stochastic protocols.

\subsection{Weak measurements and quantum trajectories}
Measurement lies at the core of quantum mechanics. Given a state $|\psi\rangle$, and an observable $O$, the von Neumann postulate states that, when a measurement takes place, the wavefunction collapses in the eigenvector $|n\rangle$ of $O$ corresponding to the measurement outcome $o_n$. This can be seen as the transformation
\begin{align}
	|\psi\rangle \mapsto |\psi'\rangle = \frac{P_n |\psi\rangle}{||P_n |\psi\rangle||}.
\end{align}
Here $P_n=|n\rangle\langle n|$ is the projector to the eigenspace of $o_n$ (the observable is resolved as $O=\sum_n o_n P_n$), and $p_n\equiv \langle \psi| P_n |\psi \rangle$ is the measurement probability.

These projective measurements are an idealization: they are implicitly instantaneous, and they collapse a pure state into a pure state. In contrast, real world experiments are imperfect: they require a finite time of interaction with the system, and in general the outcome state is mixed. 
Positive operator-value measurements represent a generalization of projective measurements, which are suitable to treat situations where the interaction time between system and ancilla is comparable to the energy-scale of the system.

Given a set of positive operators $E_n$ which sum to the identity $\sum_n E_n = 1$, it is possible to define the probabilities $p_n=\mathrm{tr}(\rho E_n)$. Differently from the projective case, these operators are not orthogonal, hence the knowledge of $E_n$ does not fix the state after the measurement. It is necessary to know a set of operators $A_{n,k}$ such that $E_n = \sum_k A^\dagger_{nk} A_{nk}$ (dubbed Kraus operators). 
Then, the state after the measurement outcome corresponding to $E_n$ is
\begin{align}
\label{eq:povm1}
	\rho \mapsto \rho'= \frac{\sum_k A_{nk}\rho A_{nk}^\dagger}{\mathrm{tr}(\rho E_n)}.
\end{align}
It is clear that in general, given a pure density matrix $\rho$, the outcome measurement in Eq.~\eqref{eq:povm1} is mixed. The condition to preserve purity is that $E_n = A^\dagger_{n} A_n$, \textit{i.e.}, there is only one Kraus operator per measurement. In this case Eq.~\eqref{eq:povm1} can be recast in a wave-function perspective
\begin{align}
	|\psi\rangle \mapsto |\psi'\rangle = \frac{A_n|\psi\rangle}{\sqrt{\langle \psi |E_n|\psi\rangle}}.
	\label{eq:povm2}
\end{align}
POVM can be implemented by allowing the system to interact with an ancilla for a finite time, and apply afterward a projective measurement to the ancilla. The limit when this interaction is small is the weak measurement, when typically the state of the system is mildly perturbed. 

As an example, we consider the toy model of a two-level system $|\psi\rangle=\sum_{n=0,1} c^S_n |n\rangle_S$ which interacts with a two-level ancilla initialized in a state $|a\rangle=\sum_{k=0,1} c^A_k |k\rangle_A$. 
If their interaction is infinitesimal, the unitary evolution of the combined system initialized in $|\Psi\rangle=|\psi\rangle\otimes |a\rangle$ is given by
\begin{align}
	U(\epsilon) = \exp\left[-i\epsilon \sum_j O_j^S\otimes O_j^A\right].
\end{align}
After the unitary transformation, entanglement is generated, and the state up to first order in $\epsilon$ is given by
\begin{align}
	|\Psi'\rangle(\epsilon) &= \sum_{m,n} (c^S_n c^A_m -i\epsilon  c^\mathrm{ent}_{nm}) |n\rangle_S\otimes |m\rangle_A \nonumber \\
	c_{nm}^\textup{ent} &= \sum_j \sum_{n',m'=0,1} O^S_{j,n',n} O^A_{j,m',m} c^S_{n'} c^A_{m'}\nonumber\\
	O^\mu_{j,k,k'} &\equiv \langle k| O^\mu |k'\rangle_\mu,\quad \mu=S,A.\label{eq:povm3}
\end{align}
Rearranging the above equation, the system is given by
\begin{align}
	|\Psi'\rangle(\epsilon) = \alpha(\epsilon) |\phi_0\rangle_S\otimes |0\rangle_A + \beta(\epsilon) |\phi_1\rangle_S\otimes |1\rangle_A,
	\label{eq:povm4}
\end{align}
where the constants $\alpha,\beta$ and $\phi_k$ are read out matching Eq.~\eqref{eq:povm3} and Eq.~\eqref{eq:povm4}. If a measurement is now performed on the ancilla along the $z$-direction, one obtains
\begin{align}
	|\psi\rangle =\begin{cases}
		 |\phi_0\rangle,&  \text{with probability }\quad p_0 = |\alpha(\epsilon)|^2,\\
		 |\phi_1\rangle,&  \text{with probability }\quad p_1 = |\beta(\epsilon)|^2.
	\end{cases}
\end{align}
Hence, POVM can be implemented by carefully designing the system-ancilla interaction, the ancilla initial state, and the projective measurement on the ancilla. 
To conclude, let us notice that, in the case of two-level ancilla, the Kraus operator are one for each measurement type. Hence, the measurements are generated by $E_k = A^\dagger_k A_k$ $k=0,1$. This is a convenient, but not unique, setting to guarantee the purity-preservation of the state. Furthermore, as stressed, the precise form of the operators $A_k$ depends on the microphysics of system and ancilla.
As we shall see, to derive a stochastic Schr\"odinger equation it is usually necessary to expand up to second order in the parameter $\epsilon$. (Nonetheless, the above argument is still valid. )
In the following we shall assume that the Kraus operators (and hence the POVM) are given by the problem, thus simplifying the discussion. 

\subsection{Quantum state diffusion}
We consider the following Kraus operators
\begin{align}
	A_0 &= \sqrt{\frac{1}{2}} |0\rangle\langle 0| + \sqrt{\frac{1}{2}-\epsilon}|1\rangle\langle 1|\\
	A_1 &= \sqrt{\frac{1}{2}} |0\rangle\langle 0| + \sqrt{\frac{1}{2}+\epsilon}|1\rangle\langle 1|.
\end{align}
Starting from the state $|\psi\rangle = \alpha |0\rangle + \beta|1\rangle$, after the measurement the state is in one of the following two states
\begin{align}
	|\phi_0\rangle &= \frac{A_0 |\psi\rangle}{\sqrt{p_0}}  = \alpha\left(|\beta|^2 \epsilon + \frac{3}{2} |\beta|^4 \epsilon^2\right)|0\rangle \nonumber \\
	& +\beta \left(\epsilon (1-|\beta|^2) + \frac{\epsilon^2}{2}(-1-2|\beta|^2+3|\beta|^4)\right)|1\rangle\\
	|\phi_1\rangle &= \frac{A_1 |\psi\rangle}{\sqrt{p_1}}  = \alpha\left(-|\beta|^2 \epsilon + \frac{3}{2} |\beta|^4 \epsilon^2\right)|0\rangle \nonumber \\
	& +\beta \left(-\epsilon (1-|\beta|^2) + \frac{\epsilon^2}{2}(-1-2|\beta|^2+3|\beta|^4)\right)|1\rangle,
\end{align}
with probability respectively $p_0 = 1/2 -|\beta|^2 \epsilon$ and $p_1= 1/2 + |\beta|^2\epsilon $. The above post-measurement state can be collected into a  compact differential form, by introducing the random binomial variable $dW=\mp \epsilon $,  and the operator $n = |1\rangle\langle 1|$. We have
\begin{align}
	d|\psi\rangle &= |\psi'\rangle - |\psi\rangle = 2 \epsilon^2 (\langle n\rangle^2 -2 \langle n\rangle n)|\psi\rangle \nonumber\\
	&+dW (n-\langle n\rangle) |\psi\rangle - \frac{\epsilon^2}{2} (n-\langle n\rangle)^2|\psi\rangle,
\end{align}
where $\langle \circ \rangle = \langle \psi |\circ|\psi\rangle$ is the state average.
The expression can be simplified by centering the random variable $dW\mapsto d\xi = dW-\overline{dW}$, since the average $\overline{dW}=2\beta^2\epsilon^2= 2\langle L\rangle \epsilon^2$ due to the $\mathcal{O}(\epsilon)$ unbalance in the outcome probability. The variance is preserved at leading order $d\xi^2 = \epsilon = dW^2$, and the equation reads
\begin{align}
	d|\psi\rangle &= d\xi (n-\langle n\rangle) |\psi\rangle - \frac{\epsilon^2}{2} (n-\langle n\rangle)^2|\psi\rangle.
	\label{eq:povm6}
\end{align}

Eq.~\eqref{eq:povm6} gives one interaction with a single ancilla qubit. In order to derive a stochastic Schr\"odinger equation we subsequently couple a series of identically prepared ancillas with the same paradigm. Hence, Eq.~\eqref{eq:povm6} is iterated multiple time. Taking the scaling limit, with infinitesimal time step we have
\begin{align}
	\frac{d|\psi(\xi_t)\rangle}{dt}&= \frac{d\xi_t}{dt} (n-\langle n\rangle_t) |\psi(\xi_t)\rangle \nonumber \\&\qquad - \frac{\epsilon^2}{2 dt} (n-\langle n\rangle_t)^2|\psi(\xi_t)\rangle.
\end{align}
Fixing the scaling $\epsilon^2 = \gamma dt$ reduce this equation to
\begin{align}
	d|\psi(\xi_t)\rangle &= d\xi_t(n-\langle n\rangle_t) |\psi(\xi_t)\rangle \nonumber \\
	&\qquad- \frac{\gamma}{2}dt (n-\langle n\rangle_t)^2|\psi(\xi_t)\rangle.
\end{align}
This is the quantum state diffusion equation for a two-level qubit. 

Let us make some final comments. (i) The scaling $\epsilon =\sqrt{\gamma dt}$ is typical when dealing with Kraus operators (for instance, see  Ref.~[\onlinecite{plenio1998the,brun2002a,breuer2002the}] for an indepth discussion). This is due to a Markovian approximation, which is assumed to hold for the systems of interest. (ii) The generalization to multiple qubits is trivial. Recovering the noise terms in Eq.~\eqref{eq:sse} require the introduction of $L$ uncorrelated \^Ito terms $d\xi_i$ with $\overline{d\xi} = 0$ and $d\xi_i d\xi_j=\gamma dt \delta_{ij}$. (iii) For the time evolution in the limit $dt\to 0$, the binomial variable can be approximated by a Gaussian increment, in the same spirit as obtaining a Wiener process from a random walk. 

\subsection{Quantum jumps and non-Hermitian Hamiltonian}
We consider the following Kraus operators
\begin{align}
	A_0 &=  |0\rangle\langle 0| + \cos\epsilon |1\rangle\langle 1|\\
	A_1 &=  \sin{\epsilon} |1\rangle\langle 1|.
\end{align}
In this situation, given the state $|\psi\rangle = \alpha|0\rangle + \beta|1\rangle$, the measurements $E_0$ and $E_1$ are unbalanced, with probability $p_0 = 1-|\beta|^2\epsilon^2$ and $p_1=|\beta|^2\epsilon^2$, respectively.
As in the previous section, we introduce the operator $n=|1\rangle\langle 1|$. The state after the measurement is in either of the following states
\begin{align}
	|\phi_0\rangle &= |\psi\rangle + \frac{1}{2} \alpha \langle n\rangle \epsilon^2 |0\rangle + \frac{1}{2}\beta (\langle n\rangle -1)\epsilon^2|1\rangle\nonumber \\
	& = |\psi\rangle-\frac{1}{2}\left(n-\langle n\rangle\right)\epsilon^2|\psi\\
	|\phi_1\rangle & = |1\rangle.
\end{align}
In differential form, we have
\begin{align}
	d|\psi\rangle = -\frac{1}{2}\left(n-\langle n\rangle\right)\epsilon^2|\psi\rangle + \left(\frac{n}{\sqrt{\langle n\rangle}}-1\right)\delta N|\psi\rangle.
	\label{eq:povm8}
\end{align}
Here $\delta N$ is a Poisson process: $\delta N=0(1)$ with probability $p_0(p_1)$. Since $\overline{\delta N} = |\beta|^2 \epsilon^2$, the state is unlikely to experience a measurement and usually is affected only by the Kraus operator $A_0$. However, when $\delta N=1$, the second term in Eq.~\eqref{eq:povm8} dominate the infinitesimal one, and project the state onto $|1\rangle$. Notice that in this case $\delta N^2=\delta N$. 

Suppose that the system interacts with $n$ ancilla qubits. As previously discussed, the stochastic Schr\"odinger equation is given once the scaling $\epsilon^2=\gamma dt$ is set
\begin{align}
	d|\psi(N_t)\rangle &= -\frac{1}{2}\left(n-\langle n\rangle_t \right)\gamma dt|\psi(N_t)\rangle\nonumber \\
	&\qquad + \left(\frac{n}{\sqrt{\langle n\rangle_t}}-1\right)\delta N_t|\psi(N_t)\rangle.
	\label{eq:povm9}
\end{align}
This is the so-called quantum jump equation. The first term is a non-Hermitian Hamiltonian, while the second term is the quantum jump. When the event happens, the state is projected to $|1\rangle\langle 1|$. 
It is interesting to ask how much is the probability that the state does never jump? The probability of this no-click dynamics for one-qubit is given by
\begin{align}
	P_\textup{no-click} = \Pi_{n=0}^\infty p_0 \simeq |\alpha|^2.
\end{align}
Thus, the number of trajectories with no-jump effect is substantial.

We conclude with few remarks. (i) The generalization to many-body is trivial, once $L$ uncorrelated variables are included, and the Hamiltonian $H$ of the system is added. The final equation for the quantum jumps is Eq.~\eqref{eq:qjump}. The non-Hermitian Hamiltonian in Eq.~\eqref{eq:nonherm} is recasted when post-selecting the trajectories without jumps. Notice that the contribution $\propto \gamma dt \sum_i \langle n_i\rangle_t$ factorize as an overall rescaling of the wavefunction, and cancels out in the average.
(ii) In the situation of a spin-chain, the probability of no-click dynamics (hence non-Hermitian generated dynamics) depends on the system size. In practice, one chooses among all the trajectories the portion where no jump has occurred. This post-selection process is key in investigating non-Hermitian physics in experimental setups. 
(ii) Both the quantum jump and the quantum state diffusion equations induce the same Lindblad evolution Eq.~\eqref{eq:lind} for the mean state. This difference highlights how quantum trajectories contain substantial more information than the average state. Importantly, the non-Hermitian Hamiltonian coincide with the non-coherent term of the Lindblad equation~\eqref{eq:lind}.

\section{Free fermion techniques and numerical implementation}
\label{app:ff}
In this section we present a summary of the simulation techniques used in this paper. (For a general reference on the ideas used here, see Ref.~[\onlinecite{mbeng2020the}]).
These methods are based on the fermionic representation of the Ising model, which can be applied also to Eq.~\eqref{eq:sse} (Eq.~\eqref{eq:nonherm}), as the noise (non-Hermitian) term preserve the quadratic structure. We first review the Jordan-Wigner transformation, and explain how the time-evolution reads in fermionic variables. Due to the Gaussian preservation of both quantum state diffusion and non-Hermitian Hamiltonian, the state is fully characterized by its two point function. Afterwards, we show how the entanglement entropy is derived from the correlation matrix. Finally, we give some detail of the numerical implementation. 

\subsection{Jordan-Wigner mapping}
The Jordan-Wigner transformation is defined by
\begin{align}
\label{eq:jwv0}
	\sigma^x_i &= K_i (c_i+ c^\dagger_i),\quad K_i = \prod_{j<i} (2 n_i -1)\\
	n_i & = c_i^\dagger c_i.
\end{align}
Here, the string operator $K_i$ ensure the fermionic algebra. For the quantum state diffusion Eq.~\eqref{eq:sse}, the dynamics maps to
\begin{align}
	d|\psi(\vec{\xi}_t)\rangle &= -i \sum_{i=1}^{L-1} (c^\dagger_i c_{i+1} + c^\dagger_i c^\dagger_{i+1} + \mathrm{h.c.})dt |\psi(\vec{\xi}_t)\rangle \nonumber \\
	& \quad + \sum_{i=1}^L (n_i-\langle n_i\rangle_t) d\xi^i_t |\psi(\vec{\xi}_t)\rangle \nonumber \\
	& \qquad - \frac{\gamma}{2}dt\sum_{i=1}^L (n_i-\langle n_i\rangle_t)^2  |\psi(\vec{\xi}_t)\rangle.
	\label{eq:sseffv0}
\end{align}
Instead for the non-Hermitian Hamiltonian Eq.~\eqref{eq:nonherm}, we have
\begin{align}
	d|\psi\rangle &= -i \sum_{i=1}^{L-1} (c^\dagger_i c_{i+1} + c^\dagger_i c^\dagger_{i+1} + \mathrm{h.c.})dt |\psi\rangle \nonumber \\
	& \qquad - \frac{\gamma}{2}dt\sum_{i=1}^L n_i  |\psi\rangle.
	\label{eq:nonhermv0}
\end{align}
Within this formalism, if the initial state of the system is Gaussian, then the state is Gaussian at each time step. In particular, all the observables are encoded in the correlation matrix
\begin{align}
	G_{ij} = \begin{pmatrix}
		\langle c^\dagger_i c_{j}\rangle & \langle c^\dagger_i c_{j}^\dagger\rangle\\
		\langle c_i c_{j}\rangle & \langle c_i c^\dagger_{j}\rangle.
	\end{pmatrix}\label{eq:corrmat}
\end{align}
In the next subsection, we show how the dynamics can be reduced to the time-evolution of a $2L\times 2L$ matrix. 

\subsection{Dynamics in the fermionic formalism}
The equation of motion for both the protocol can be written as $d|\psi(\vec{\xi}_t)\rangle = dZ_t |\psi(\vec{\xi}_t)\rangle$, where $dZ_t$ is quadratic in the creation/destruction fields. For the quantum state diffusion
\begin{align}
	dZ_t &= dt\left(-i H  -\frac{\gamma}{2}\sum_i (n_i-\langle n_i\rangle_t)^2\right) 
	\nonumber \\&\qquad + \sum_i d\xi^i_t (n_i-\langle n_i\rangle_t),
\end{align}
while for the non-Hermitian Hamiltonian we have
\begin{align}
	dZ_t = dt\left(-i H  -\frac{\gamma}{2}\sum_i n_i\right) = -i dt H_\mathrm{eff}.
\end{align}
Both these instances can be treated on the same foot, once the operator $dZ_t$ is specified. From the operator $dZ_t$, it is possible to write down an equation of motion for the correlation matrix Eq.~\eqref{eq:corrmat}. (This is obtained by taking the differential of Eq.~\eqref{eq:corrmat}, simplifying the fermionic algebra, and expanding up to order $\mathcal{O}(dt)$). 

However, in practical implementation, this method is not the most efficient. A better strategy, which we follow in this paper, is the route pioneered in Ref.~[\onlinecite{cao2019entanglement}]. Since the system is Gaussian, for any $t$, there exist a unitary matrix
\begin{align}
	\mathcal{U}_t = \begin{pmatrix}
		U_t & V^\dagger_t\\
		V_t & U^\dagger_t,
	\end{pmatrix}
\end{align}
such that
\begin{align}
	|\psi\rangle_t = \frac{1}{\sqrt{|\det U_t|}} \exp\left(-\frac{1}{2} \sum_{i,j,k} (U_t)_{i,k}(V_t^\dagger)_{k,j} c^\dagger_i c^\dagger_j \right)|0\rangle.
	\label{eq:state}
\end{align}
This condition is equivalent to finding a unitary transformation that maps the original fermionic fields $(c_i,c_j^\dagger)$ to a new set $(\chi_i,\chi_j^\dagger)$ such that $\chi_i|\psi\rangle_t=0$ for all $i$. Explicitly, labelling a vector of fermionic field as $\mathbf{c} = (c_1,c_2,\dots, c_L)$,
\begin{align}
	\begin{pmatrix}
		\mathbf{\chi}\\
		\mathbf{\chi}^\dagger
	\end{pmatrix} = \mathcal{U}_t \begin{pmatrix}
		\mathbf{c}\\
		\mathbf{c}^\dagger.
	\end{pmatrix}
	\label{eq:heis}
\end{align}
From the definition of the correlation matrix and Eq.~\eqref{eq:state}, and after few manipulation in fermionic algebra, we have the relationship
\begin{align}
	G(t) = \mathcal{U}_t \begin{pmatrix}
		 \mathbf{1}_{L\times L}& \mathbf{0}_{L\times L}\\
		\mathbf{0}_{L\times L} & \mathbf{0}_{L\times L}	\end{pmatrix}\mathcal{U}^\dagger_t.\label{eq:correlazione}
\end{align}
Hence, the full evolution is encoded in $\mathcal{U}_t$. The equation of motion for this object are easily derived from Eq.~\eqref{eq:heis} and using the annihilation condition  $\chi_i|\psi\rangle_t=0$. In practice it can be implemented into two steps. (i) Integrate the Heisenberg equation of motion between $t$ and $t+dt$ (with initial condition $\mathcal{U}_{t=0} = \mathcal{U}_0$ given)
\begin{align}
	d \mathcal{U}_t = 2 d\mathcal{Z}_t \mathcal{U}_t.
\end{align}
(ii) Apply a renormalization to guarantee that $U_{t+dt}$ is unitary (for instance a QR decomposition). This guarantees that $|\psi\rangle_{t+dt}$ is in the form Eq.~\eqref{eq:state} with the operators given in $U_{t+dt}$.
For concrete details on the numerical implementation, see Sec.~\ref{subsec:numimpl}.

\subsection{Majorana fermions and entanglement entropy}
\label{sec:majoent}
For a Gaussian state, Wick's theorem reduce the computation of entanglement from diagonalizing a $2^L\times 2^L$ matrix to a $2 L\times 2 L$ linear problem. For this goals, it is convenient to introduce two species of Majorana fermions
\begin{align}
	a_{j,1} = c_j + c_j^\dagger,\qquad  a_{j,2} = i (c^\dagger_j-c_j).
\end{align}
Then the matrix
\begin{align}
	V = \begin{pmatrix}
		\mathbf{1}_{L\times L} & \mathbf{1}_{L\times L}\\
		-i \mathbf{1}_{L\times L} & i \mathbf{1}_{L\times L } 
	\end{pmatrix},
\end{align}
transforms the correlation function from Dirac fermions to Majorana $W = V G V^\dagger$. Due to the Majorana algebra $\{a_{i,p},a_{j,q}\}= \delta_{i,j}\delta_{p,q}$, the Majorana correlation can be decomposed into $W=\mathbf{1}+i \tilde{W}$. The matrix $\tilde{W}$ is real and anti-symmetric, and is the key actor in computing any observable. In particular, entanglement entropy on a bipartition $A\cup B$ is obtained by diagonalizing $\tilde{W}|_{A(B)}$, that is restricting the indices of the Majorana fermions within either partition. 
If the eigenvalues of $\tilde{W}|_A$ are $\lambda_k$, then the entanglement entropy is given by
\begin{align}
	S &= -\sum_k \left(\frac{1+\lambda_k}{2}\ln \frac{1+\lambda_k}{2} +\frac{1-\lambda_k}{2}\ln \frac{1-\lambda_k}{2}\right).
\end{align}

\subsection{Numerical implementation}
\label{subsec:numimpl}
The numerical implementation can be adapted to the protocol considered. For the quantum state diffusion, it is practical to divide the unitary step from the noise term $d\mathcal{Z}_t = -i\mathcal{H} dt + dT$. 
The Hamiltonian $H$ can be stored and exponentiated only once, while the noise term $dT$ is diagonal, with elements $dT_{ij} =\delta_{ij}(d\xi^i+\gamma dt (2\langle n_i\rangle-1))/2$ that needs to be computed repeatedly at each time step.

The resulting evolution then is
\begin{align}
	\tilde{\mathcal{U}}_{t+dt} =\mathcal{N} e^{2dT} e^{2i \mathcal{H}dt}\mathcal{U}_{t},
	\label{eq:preqr}
\end{align} 
where we have neglected higher order terms $\mathcal{O}(d\xi dt)$ (Trotter approximation). The normalization $\mathcal{N}$ is unimportant, as Eq.~\eqref{eq:preqr} is followed by a QR decomposition $\tilde{\mathcal{U}}_{t+dt} = \mathcal{Q}_{t+dt}\mathcal{R}_{t+dt}$, and by setting $\mathcal{U}_{t+dt} = \mathcal{Q}_{t+dt}$.
As previously discussed, this condition guarantees the state is in the form Eq.~\eqref{eq:state} and the correlation function is obtained through Eq.~\eqref{eq:correlazione}.

For the non-Hermitian quantum evolution, it is sufficient to exponentiate once the   operator $d\mathcal{Z}_t = -i \mathcal{H}_\mathrm{eff} dt$ and apply it subsequently to generate the evolution
\begin{align}
	\tilde{\mathcal{U}}_{t+dt} =\mathcal{N}e^{2i \mathcal{H}_\mathrm{eff}dt}\mathcal{U}_{t}.
\end{align} 
After each time step, the QR decomposition is implemented to normalize the state.
In both situations, the entanglement entropy is computed as described in Sec.~\ref{sec:majoent}.


\end{document}